\documentclass[final]{dmtcs-episciences}

\usepackage[utf8]{inputenc}
\usepackage{subfigure}

\usepackage[round]{natbib}
\usepackage{amsmath,amssymb}
\usepackage{bm} %OUR ADDITION
\usepackage{appendix} %OUR ADDITION
\usepackage{xcolor}
\usepackage{multirow} %Added
\author{Raheel Anwar\affiliationmark{1}
  \and Muhammad Irfan Yousuf\affiliationmark{2}
  \and Muhammad Abid\affiliationmark{3}}
\title[Generating Weakly Scale-free Networks]{Analysis of a Model for Generating Weakly Scale-free Networks}
\affiliation{
  Karl Franzens Universität,Graz, Austria\\
  Korea Institute of Science and Technology, Seoul, Korea\\
  Pakistan Institute of Engineering and Applied Sciences, Islamabad, Pakistan}
\received{2019-9-27}
\revised{2020-1-10}
\accepted{2020-1-13}
\begin{document}
\publicationdetails{22}{2020}{1}{7}{5791}
\maketitle
\begin{abstract}
  It is commonly believed that real networks are scale-free and the fraction of nodes $P(k)$ with degree $k$ satisfies the power-law $P(k) \propto k^{-\gamma}$  for $k > k_{min} > 0$. Preferential attachment is the mechanism that has been considered responsible for such organization of these networks. In many real networks, degree distribution before the $k_{min}$ varies very slowly to the extent of being uniform as compared to the degree distribution for $k > k_{min}$. In this paper, we propose a model that describes this particular degree distribution for the whole range of $k>0$. We adopt a two step approach. In the first step, at every time stamp we add a new node to the network and attach it to an existing node using preferential attachment method. In the second step, we add edges between existing pairs of nodes with the node selection based on the uniform probability distribution. Our approach generates weakly scale-free networks that closely follow the degree distribution of real-world networks. We perform a comprehensive mathematical analysis of the model in the discrete domain and compare the degree distribution generated by this model with that of real-world networks.
\end{abstract}

\section{Introduction} \label{sec:in}
Complex Networks have a wide variety of applications in many domains including economics, business, transportation, and natural, social and computer sciences, to name a few. Examples of such networks include online social networks, biological networks, technological networks, scientific collaboration networks, citation networks and many more. The complex structure of these networks has remained the main focus of studies in the past. Many studies claim that these networks are scale-free networks. However, contrary to this common belief, recent studies show that scale-freeness is rare in real-world networks \cite{SF_rare}. Generally, a network is scale-free if the fraction of nodes with degree $k$ follows a power-law distribution $k^{-\gamma}$, where $\gamma >$ 1. Typically, it is believed that the $\gamma \in [2,3]$ or in other words the degree distribution of such networks follows a preferential attachment, i.e., the rich-get-richer mechanism. In order to understand these networks in detail, many generative models have been proposed to artificially generate such networks. In a recent study \cite{SF_rare} the authors test the universality of the scale-free structure by applying different statistical tools to a large corpus of nearly 1000 network data sets drawn from social, biological, technological, and informational sources. They observe that only 4\% of real-world networks show strong evidence of scale-free structure. Furthermore, they find that social networks are at best weakly scale-free, while only a handful of technological and biological networks can be called strongly scale-free.

There are several studies, e.g., \cite{Barabasi,PA1,PA2,PA3, PA4, PA6} to generate scale-free networks such that the value of $\gamma \in [2,3]$. However, to the best of our knowledge, there has been no work to generate weakly scale-free networks. We define a weakly scale-free network as a network that has the power-law exponent $\gamma >$ 3 for $k > k_{min}$ and a flattish distribution of degree of nodes for $k < k_{min}$. In this paper, we aim to generate weakly scale-free networks with a two step approach. In the first step, named node-step,  a new node enters the network and connects to an existing node using the preferential attachment method. While in the second step, named edge-step, we iteratively add a certain number of edges between existing nodes at random. This adds the desired flatness in the network before $k_{min}$ and the generated network follows the scale-free structure after $k_{min}$ with $\gamma >$ 3. Two mechanisms for creating edges are discussed. In the first mechanism, a fixed number of edges are created during the edge-step and the resulting model generates weakly scale-free network having the distribution satisfying the power-law with exponent $\gamma \in (3, \infty)$ corresponding to this fixed number in the range $[1, \infty)$. While in the second mechanism, the number of edges to be added is proportional to the fraction of the total number of existing edges by the number of nodes, we get $m(t) \propto t^\beta $ (see Appendix \ref{Appendix B}). Moreover, we choose $\beta > 1$ to further investigate network evolution when a large number of edges are added at random in each time step. The growth of the resultant network is accelerated growth.  We have found a non-stationary degree distribution for this case. The average degree, in this case, is proportional to $t^{\beta-1}$ and grows over time $t$.

For each method of adding edges, we have performed a fully detailed analysis in the discrete domain ($k$ and $t$ as discrete variables). During the edge-step, edges are added iteratively and we update the number of nodes marked by degree $k$ after each iteration. A detailed formulation for each step derived in this paper depicts the process fully. An exact recurrence relation is formulated that models the network transition from one state to another. It describes the change in degree distribution from the node-step to the edge-step. It also models the transition from the state at the previous time stamp to the next time stamp when both the node- and edge- steps have been accomplished.

The idea of adding edges between existing nodes for the network evolution is not a new one. The authors in the paper \cite{D1} extended the Barabasi-Albert (BA) model and presented the undirected network evolution based on this idea. They proposed the model in which a new incoming node preferentially connects to $m$ existing nodes of the network. Then, simultaneously $cm, c>0$ edges appear between existing nodes. The selection of nodes for these new edges is also made preferentially. For a new edge, a pair of nodes is selected with the probability proportional to the product of their degrees. They used the continuous approximation (considering $t$ and $k$ as continuous variables, \cite{D4}) and found that the degree-distribution satisfies the power-law $k^{-\gamma}$ with $\gamma = 2 + \frac{1}{1+2c}$. This approach was proved to describe quite well the degree distributions of networks growing under the mechanism of preferential linking. \cite{D4}) discussed another method of growing directed networks through mixing preferential and random linking. A new vertex with $n$ incoming edges enters the network, at each time step. The target ends of $m$ new edges are simultaneously distributed among vertices through the preferential linking rule. The probability to select a node is proportional to $q+A$, with $q$ is the in-degree of the nodes and $A$ is a constant with value $A > -n - n_r$. Besides, the target ends of $n_r$ new edges are attached to randomly chosen vertices with source ends of each edge may be anywhere. The in-degree $P(q)$ of these nodes satisfies the relation $q^{-\gamma}$ with $\gamma = 2 + \frac{n_r + n + A}{m}$. These were kind of non-equilibrium growing networks. The average degree for networks having the exponent $\gamma > 2$ tends to some constant value. The total degree of the network is a linear function of the number of nodes at time $t$. When the average degree does not approach a constant value but grows over time(total degree grows at least super-linearly), these networks are called networks with accelerated growth\cite{D2}. In general, the total number of edges in networks with accelerated growth is $m(t) \propto t^\beta$ for $\beta > 1$. In \cite{D3} authors present a network with acceleration growth in which $ct$ number of edges emerge among old vertices at each time step. This behavior is found in the Internet, collaboration networks, the World Wide Web, and many other networks.

The rest of the paper is organized as follows. In section 2, we present a two-step model for generating weakly scale-free networks. In section 3, we provide a comprehensive mathematical analysis of our model and drive recurrence relation for both the node- and edge- steps. In section 4, we compare the graphs generated by our model with real-world networks. In section 5, we discuss related work and conclude the paper in section 6.

\section{Weakly Scale-free Model} \label{sec:first}

In this section, we formally present our model called Weakly Scale-free Model (WSM). WSM generates undirected graphs and can be run for any period of time $t (t > 0)$, given an initiator graph at $t = 0$, that could be any connected graph, e.g., a connected triplet. We represent a network as an undirected graph $G = (V, E)$ where $V$ is the set of nodes and $E$ is the set of edges. We represent nodes in the graph as $\{v_1,v_2,v_3, ..., v_n \}$ and edges as $\{e_1, e_2, e_3, ..., e_m\}$, furthermore, $|V|=n$ and $|E|=m$. In this model, a network evolves over time and at any instant of time $t$, there are $|V_t|=n_t$ nodes and $|E_t|=m_t$ edges in the network. Let $G_t (V_t , E_t )$ be the graph at time $t$ .  In a single time stamp, we perform two steps namely node-step and edge-step.\\
\textbf{Node Step:} A new node $x$ arrives and connects to an existing node $y$ via one edge such that
\begin{equation}
n_{t+1} = n_t + 1
\end{equation}
we call it the node equation. The probability $P_y$ of selecting node $y$ is proportional to its degree $d_{y}$ i.e.,		
\begin{equation}
P_y = \frac{d_y}{\sum_{i=1}^{n} d_i}
\end{equation}
This equation suggests that a high degree node attracts more nodes and helps the network to follow a power-law for its degree distribution.\\
\textbf{Edge Step:} In the edge-step, we add $\Delta m_t$ number of edges in the network at each time stamp such that
\begin{equation}
m_{t+1} = {m_t} + \Delta m_t +1
\end{equation}
We call it the edge equation. Here, $+1$ corresponds to the edge added in the node-step. We add edges between existing nodes at random. In order to add an edge, we select two nodes $x$ and $y$ uniformly at random from the graph $G_t$ and adds an edge between them. we repeat this process $\Delta m_t$ times. Here, $\Delta m_t$ is a natural number that can have any positive values, $\Delta m_t \in [0,1,2,3, ...]$. In other words, we add a {fixed number of edges} in the edge-step. However, a formulation for the variable number of edges is also presented in the appendix. To account for both of these two cases, we used a more generic term $\Delta m_t$ to represent the number of added edges during the edge-step. When $\Delta m_t=0$, our model generates a standard Barabasi-Albert network and when $\Delta m_t > 0$, we add random links in the network and as a result we generate weakly scale-free networks.

\section{Mathematical Analysis}
In this section, we present the mathematical analysis of WSM in detail. For analysis, let's assume that the network starts with a single node at time $t=1$. It grows over time through two steps, namely the node-step and edge-step as described above. During the edge-step, $\Delta m_t$ number of edges are created. For the creation of each edge, two nodes from the network are selected according to the uniform probability distribution function. A recurrence relation for degree distribution is formulated that incorporates both node and edge steps and describes overall network evolution through time.

More formally, let $\{v_1, v_2, v_3, ..., v_n\}$ represents nodes in the network and $\{e_1, e_2, e_3, ..., e_m\}$ edges. The network evolves from a single node at time $t=1$. For ease of analysis,  time $t$ also refers to the total number of nodes in the network, i.e., at time $t$ there are a total of $t$ nodes in the network. The degree of a node is represented as $k$. The network evolves in two steps. In the first step, a new node is added that connects with an existing node via one edge. In the second step, we add an edge between two randomly selected nodes. This step is repeated until $\Delta m_t$ new edges are created in the network. For creating edges, we consider a simple scenario in which $\Delta m_t$  is a fixed positive integer $\alpha$. Formulation keeps a record of the number of nodes of degree $k$  at each time step $t$. $N_{k,t}$ is the number of nodes in a group marked by degree $k$ at the completion of time step $t$. After the node-step, wherein $(t+1)$th node connects to the network, number of nodes in the group becomes $N_{k,t+1,0}$. $N_{k,t+1,l}$ represents nodes count of $k-th$ degree during the edge-step. The subscript $l \in [1,\Delta m_t]$ tells that $l$ number of edges have been added during the edge-step. The network state at the completion of both steps is represented by $N_{k,t+1, \Delta m_t}$, which is also the initial configuration $N_{k,t+1}$ for the next time step. The same process is repeated for the next time step. To analyze the network behavior over time, the rank of nodes marked by degree $k$, define as $P_{k,t} = \frac{N_{k,t}}{t}$, is more helpful than just keeping a record of actual count of these nodes. $P_{k,t}$ represents fraction of nodes of degree $k$.
Briefly, two consecutive stages through which network grows, for each time step,

\begin{enumerate}
	\item (Node-step): At time $t+1$, $(t+1)$-th node  arrives and connects to an member node via one edge. The older node is picked with probability proportional to its degree $k$.
	\item (Edge-step): At time $(t+1)$, after the node-step, two nodes are selected randomly (with uniform probability) and an edge is created between them. This step is repeated $\Delta m_t$ number of times. $\Delta m_t$ edges are added during the edge-step.
\end{enumerate}

First, we describe the model that incorporates the node-step only, and later we extend it to include the edge-step.

\subsection{The Standard BA Model}
Our network is an undirected graph. In an undirected graph, $\sum_{n=1}^{t}k_{n} = 2m_t$ where $k_n$ is the degree of $n$-th node and $m_t$ is the total number of edges in the graph at the completion of time step $t$. In the case where network evolution does not include the edge-step, it is described by the BA model. The network has $t-1$ edges at time step $t$.

The standard BA model gives following recurrence relation for the degree distribution,

\begin{equation} \label{intro2}
P_{1} = \frac{2}{3}
\end{equation}

and
\begin{equation} \label{intro1}
P_{k} =  P_{k-1}.\frac{k-1}{k+2}
\end{equation}

In a closed form,

\begin{equation}\label{intro3}
P_{k} = \frac{4}{k(k+1)(k+2)} \approx \frac{c}{k^3}  ~(\text{ for large } k \text{'s})
\end{equation}

\subsection{Edge Equation}

In our work, the network configuration changes through two steps. First, through the node-step when a new node is added to the network as described above. Secondly, through the edge-step, when new edges are added between existing pairs of nodes in the network. During the edge-step, we select two nodes at random with uniform probability from the network and create an edge between them. This step is repeated to add a total of $\Delta m_t$ number of edges at the time step $t+1$.
Following events may occur during the edge creation:
\begin{enumerate}
	\item $E^{k,t+1,l}_{1}$ is the event that exactly one node is selected from the group marked by the degree $k$. $p(E^{k,t+1,l}_{1})$ is the probability of this event. The occurrence of this event reduces the population of nodes of degree $k$ by 1.  \\
	\item $E^{k,t+1,l}_{2}$ is the event that exactly two  node are selected  from the group of nodes marked by the degree $k$. $p(E^{k,t+1,l}_{2})$ is the probability of this event. The occurrence of this event reduces the population of nodes of degree $k$ by 2.  \\
	\item $E^{k-1,t+1,l}_{1}$ is the event that exactly one  node is selected  from the group of nodes marked by the degree $k-1$. $p(E^{k-1,t+1,l}_{1})$ is the probability of this event. The occurrence of this event increases the population of nodes of degree $k$ by 1. \\
	\item $E^{k-1,t+1,l}_{2}$ is the event that exactly two  nodes are selected from the group of nodes marked by the degree $k-1$. $p(E^{k-1,t+1,l}_{2})$ is the probability of this event.The occurrence of this event increases the population of nodes of degree $k$ by 2. \\
\end{enumerate}

$N_{k,t}$ describes the network state at the completion of time step $t$. Each time step consists of the node-step and edge-step, in order. $N_{k,t}$ is also the initial configuration for the next time step $t+1$. We call $N_{k,t+1,0}$ the expected number of nodes of degree $k$ after the completion of the node-step and prior to the edge-step. The network at this stage has $t+1$ nodes. $N_{k,t+1,1}$ be the expected number of nodes of degree $k$ after adding one edge in the edge-step. The two network configurations could be related as, for $k>1$,

\begin{equation}\label{GenEdge1}
\begin{aligned}
N_{k,t+1,1} = N_{k,t+1,0} - 1.p(E^{k,t+1,0}_{1}) - 2.p(E^{k,t+1,0}_{2})+1.p(E^{k-1,t+1,0}_{1})\\
+ 2.p(E^{k-1,t+1,0}_{2}) -  0.p( E^{k-1,t+1,0}_{1} \cap E^{k,t+1,0}_{1}).
\end{aligned}
\end{equation}

And, in general,
\begin{equation}\label{GenEdgem}
\begin{aligned}
N_{k,t+1,l} = N_{k,t+1,l-1} - 1.p(E^{k,t+1,l-1}_{1}) - 2.p(E^{k,t+1,l-1}_{2})+1.p(E^{k-1,t+1,l-1}_{1})\\
+ 2.p(E^{k-1,t+1,l-1}_{2}) -  0.p( E^{k-1,t+1,l-1}_{1} \cap E^{k,t+1,l-1}_{1}).
\end{aligned}
\end{equation}

For the sake of  brevity, $N_{k,t+1,\Delta m_t}$ will be represented as $N_{k,t+1}$.

Since, during the edge-step no node enters into the group marked by the degree $1$, in the above relation (\ref{GenEdgem}) truncating terms involving $k-1$ and replacing $k$  by $1$, we get following recurrence relation for $N_{1,t+,l}$,
\begin{equation}\label{GenEdgem1}
N_{1,t+1,l} = N_{1,t+1,l-1} - 1.p(E^{1,t+1,l-1}_{1}) - 2.p(E^{1,t+1,l-1}_{2}).
\end{equation}

The total number of edges in the  network varies during the node- and edge- steps. At the completion of $t$-th time step when the $t$-th nodes is connected to the network and $\alpha$ edges have been created, total number of edges of the network are $m_t = (t-1) + \alpha(t-1)= (\alpha+1)(t-1)$. When $(t+1)$th nodes arrives and connects, the number is increased by $1$, i.e. $m_t+1$. And during the edge-step, when $l$ edges have been created, number of edges become $m_t + 1+ l$. During edge creation, each node has equal chance of selection.

\begin{equation}\label{UniformPm1}
p(E^{k,t+1,l}_{1})=2 \bigl(\frac{N_{k,t+1,l}}{t+1}\bigr) \bigl(\frac{t+1- N_{k,t+1,l}}{t}\bigr) = \frac{2(N_{k,t+1,l})(t+1- N_{k,t+1,l})}{t^2+t}
\end{equation}
and
\begin{equation}\label{UniformPm2}
p(E^{k,t+1,l}_{2})= \bigl(\frac{N_{k,t+1,l}}{t+1}\bigr) \bigl(\frac{ N_{k,t+1,l}-1}{t}\bigr)  = \frac{({ N_{k,t+1,l})}^2 - N_{k,t+1,l}}{t^2+t}
\end{equation}

This probability formula is valid for all $l$'s and for all $k$'s. Next, we formulate the edge- and node- steps. \newline

\textbf {Edge-step} \newline
For $k > 1$,\newline
Substituting probability functions given by (\ref{UniformPm1}) and (\ref{UniformPm2}) in (\ref{GenEdge1}) and simplifying the expression, we get
\[
N_{k,t+1,1} = N_{k,t+1,0} + \frac{2}{t+1}[N_{k-1,t+1,0}-N_{k,t+1,0}].
\]

This relation describes addition of the first edge, i.e. $l=1$. Rewriting it in the following form,

\[
N_{k,t+1,1} = \bigl(1-\frac{2}{t+1}\bigr)N_{k,t+1,0} + \frac{2}{t+1}N_{k-1,t+1,0},
\]

Similarly for general $l$,
\begin{equation}\label{EdgekM}
N_{k,t+1,l} = \bigl(1-\frac{2}{t+1}\bigr)N_{k,t+1,l-1} + \frac{2}{t+1}N_{k-1,t+1,l-1}.
\end{equation}

For $k = 1$
\begin{equation}\label{Edgek1}
N_{k,t+1,l} = \bigl(1-\frac{2}{t+1}\bigr)N_{k,t+1,l-1}.
\end{equation}

\textbf {Node-step} \newline
The node step is carried through preferential attachment and is given by following relationships,\newline
For $k > 1$,
\[
N_{k,t+1,0} = N_{k,t} + N_{k-1,t}\frac{k-1}{2m_t} -  N_{k,t}\frac{k}{2m_t},
\]

or re-writing in the form
\begin{equation}\label{Nodek}
N_{k,t+1,0} =\bigl(1-\frac{k}{2m_t} \bigr) N_{k,t} + N_{k-1,t}\frac{k-1}{2m_t}.
\end{equation}

And for $k =1 $
\begin{equation}\label{Nodek1}
N_{k,t+1,0} = \bigl(1-\frac{k}{2m_t} \bigr) N_{k,t} + 1
\end{equation}

\subsection{Derivation of the Recurrence relation for $\Delta m_t = \alpha$}

A general form of the recurrence relation describing the network evolution through time, incorporating both the node- and edge- steps, is presented. (See appendix \ref{Appendix A} for the detailed formulation and derivation).

For $k=1$, the recurrence relations gives $P_{k,t+1}$ in term of $P_{k,t}$,
\begin{equation}\label{Genk1}
(t+1)P_{k,t+1} = \bigl(1-\frac{2}{t+1}\bigr)^{\Delta m_t}\bigl[tP_{k,t} + 1 - \frac{k}{2m_t}tP_{k,t} \bigr].
\end{equation}

For $k>1$, the recurrence relations gives $P_{k,t+1}$ in term of $P_{k-n,t}$'s with $n=0,1,\dots $,

\begin{equation}\label{Genk}
(t+1)P_{k,t+1} = \sum_{n=0}^{q} tC_{k-n,\Delta m_t}P_{k-n,t}\\
\end{equation}

Coefficients $C_{k-n,\Delta m_t}$'s and $q$ are defined in the appendix \ref{Appendix A}.\newline

The stationary degree distribution for nodes is obtained on replacing $\Delta m_t$ by $\alpha$ and applying the limit $t \to \infty$, .\newline

\textbf{For $k=1$ }
\begin{equation}\label{UEdgeEqnM4}
P_{1} =  \frac{2(\alpha+1)}{ 4\alpha^2 + 6\alpha +3}.
\end{equation}
\newline
\textbf{In general, for $k > 1$}
\begin{equation}\label{UEdgeEqnM2}
P_{k} = \frac{k+ 4\alpha^2 + 4\alpha -1}{k + 4\alpha^2 + 6\alpha +2}P_{k-1}.
\end{equation}

$P_{k}$ can also be written in the closed form:
\[
P_{k} = \frac{(4\alpha^2 + 6\alpha +3)(4\alpha^2 + 6\alpha +2)\dots(4\alpha^2 + 4\alpha +1)}{ (k + 4\alpha^2 + 6\alpha +2)(k + 4\alpha^2 + 6\alpha +1) \dots(k + 4\alpha^2 + 4\alpha )}   \frac{2(\alpha+1)}{ 4\alpha^2 + 6\alpha +3},
\]
or
\begin{equation}\label{UEdgeEqnM3}
P_{k} = \frac{(4\alpha^2 + 6\alpha +2)\dots(4\alpha^2 + 4\alpha +1)}{ (k + 4\alpha^2 + 6\alpha +2)(k + 4\alpha^2 + 6\alpha +1) \dots(k + 4\alpha^2 + 4\alpha )}   {2(\alpha+1)}.
\end{equation}

If $\alpha$ is replaced with $0$ in the equation (\ref{UEdgeEqnM2}), equation (\ref{intro1}) is obtained which is the recurrence relation when only the node-step is used. Similarly, equation (\ref{UEdgeEqnM4}) reduces to equation (\ref{intro2}).
The degree distribution $P_k$  follows the power-law $k^{-(2\alpha+3)}$, but only for very large values of $k$. For large $\alpha$ and small $k$'s, distribution is almost uniform, i.e., $P_{k}\approx P_{k-1}$. Also, the expression for $P_{1}$ approaches to $0$ as $\alpha$ becomes large and large. To prove the claim about distribution slopes formally,  let $\delta_{\alpha,k} \text{ for } k> 1$ represents the slope of the $\log(P_k)\text{ vs. } \log(k)$ graph, then

\begin{equation}\label{SlopeMain}
\begin{aligned}
\delta_{\alpha,k} = \frac{ \log {P_k}-\log {P_{k-1}}}{\log {k}-\log {(k-1)}} = \frac{\log {(k+ 4\alpha^2 + 4\alpha -1)}- \log{(k + 4\alpha^2 + 6\alpha +2)}} {\log {k}-\log {(k-1)}}\\
=\frac {\log{( {1 - \frac{2\alpha+3}{k+4\alpha^2+6\alpha+2}})}} {\log{(1 + \frac{1}{k-1}})}
\end{aligned}
\end{equation}

For further analysis, let's expand $\log$ terms to the first order,
\[
\delta_{\alpha,k} \approx -\frac{(2\alpha+3)(k-1)}{k+4\alpha^2+6\alpha+2}.
\]
Now, as $\alpha$ is a fixed constant value, for very large $k$, say,  $k \gg 4\alpha^2+6\alpha+2$, expression becomes,

\begin{equation}\label{USlopeEqn1}
\delta_{\alpha,k} \approx -\frac{(2\alpha+3)(k-1)}{k}\approx -(2\alpha+3).
\end{equation}

For small $k$'s, we get,
\begin{equation}\label{USlopeEqn2}
\delta_{\alpha,k} \approx -\frac{(2\alpha+3)(k-1)}{4\alpha^2+6\alpha+2} \approx -\epsilon(k-1)
\end{equation}

where $\epsilon$ very small positive quantity of order $\mathcal{O} (\frac{1}{\alpha})$. Informally, in the beginning when $k$ is small, slope is very small.

\begin{figure}[h]
	\includegraphics[scale=0.6]{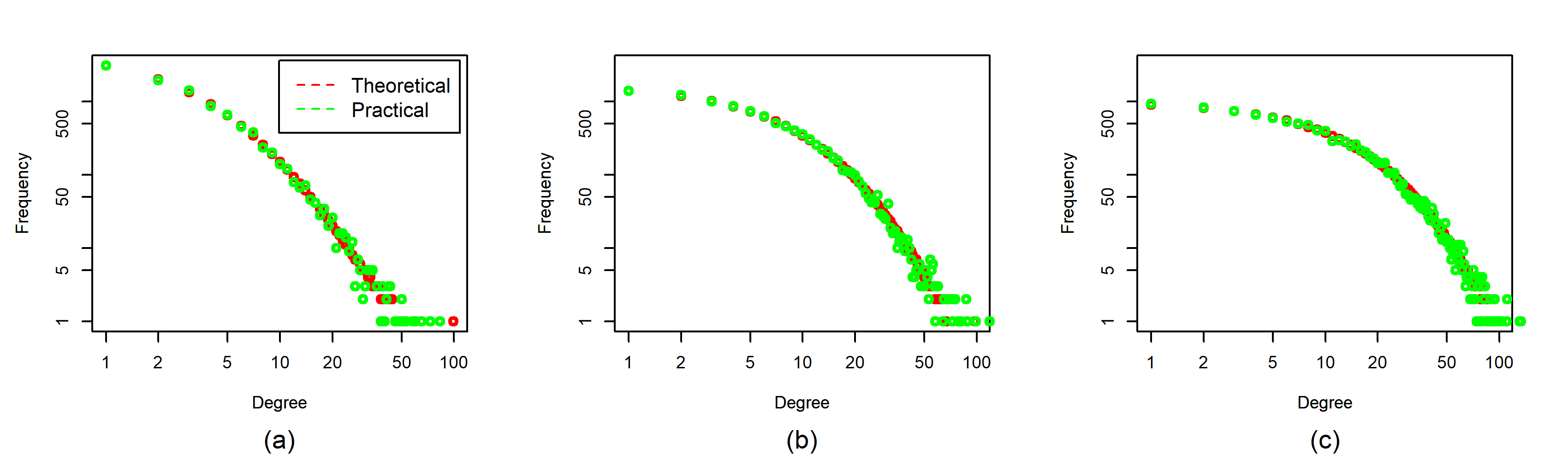}
	\centering
	\caption{Theoretical vs. Practical Degree Distribution for $\alpha=1,3,5$}
	\label{TP_fig}
\end{figure}

We compare theoretical equations and practical results in Figure~\ref{TP_fig}. For theoretical results, we implement equation ~\ref{UEdgeEqnM2} for different values of $\alpha = 1,3, 5$. For practical results, we implement the node- and edge- steps of the WS model. We draw the degree of nodes again their frequency. We see in Figure~\ref{TP_fig} that both the graphs overlap and this proves the correctness of our mathematical analysis.

\section{Experimental Results}
In this section, we provide an overview of real datasets or graphs that we will use for comparison with the networks generated by our model.  All the real networks are publicly available at \cite{konect}. We use the following six citation networks, two peer-to-peer networks and four social networks. An overview of these datasets is available in Table~\ref{Tab1}.

\begin{itemize}
	\item DBLP: This is the citation network of DBLP, a database of scientific publications such as papers and books. Each node in the network is a publication, and each edge represents a citation of a publication by another publication.
	\item HepTh: This is a citation network of authors of scientific papers from the arXiv’s High Energy Physics (Theory) section.
	\item Cora: This is the cora citation network. Nodes represent scientific papers. An edge between two nodes indicates that the one node cites the other node.
	\item HepPh: This is a citation network of scientific papers from the arXiv’s High Energy Physics (Phenomenology) section.
	\item CiteSeer: This is the citation network extracted from the CiteSeer digital library. Nodes are publications and the edges denote citations.
	\item Patents: This is the citation network of patents registered with the United States Patent and Trademark Office. Each node is a patent and an edge represents a citation.
	\item Gnutella1: This is a network of Gnutella hosts from 2002. The nodes represent Gnutella hosts, and the edges represent connections between them.
	\item Gnutella2: This is a network of Gnutella hosts from 2002. The nodes represent Gnutella hosts and the edges represent connections between them.
	\item Facebook: This is a friendship network of facebook. The nodes are users and an edge between two users shows a friendship relation between them.
	\item Prosperloans: These are loans between users of the Prosper.com website. The network denotes who loaned money to whom.
	\item Amazon: This is the network of items on Amazon that have been mentioned by Amazon's "People who bought X also bought Y" function. Nodes in the network are products, and an edge from A to B denotes that product A is frequently co-purchased with product B.
	\item Pokec: This is the friendship network from the Slovak social network Pokec. Nodes are users of Pokec and directed edges represent friendships.
	
\end{itemize}

For comparison, we compare our model with the standard Barabasi–Albert Model \cite{Barabasi}. The Barabasi–Albert (BA) Model is a method for generating scale-free networks using a preferential attachment mechanism. Since our model also applies preferential attachment mechanism so it would be interesting to compare it with the basic BA Model. In the very basic form of the  BA Model, the network begins with a small connected network of $w_0$ nodes. New nodes are added one-by-one at each time stamp and connect to $w < w_0$ existing nodes with a probability that is proportional to the number of links that the existing nodes already have. In our implementation, we start with a small connected network and connect the new nodes with $w$ number of existing nodes using the preferential attachment method described in \cite{Barabasi} where the value of $w$ is adjusted such that the resulting graph has nearly the same number of edges as that of an original graph.

\begin{table}
	\centering
	\caption {Real-world datasets used in the experiments}
	\begin{tabular}{ |p{1.8cm}|p{1.5cm}|p{1.5cm}|p{1.5cm}|}
		\hline
		
		\textbf{Datasets} &\textbf{Total Nodes} &\textbf{Total Edges} &\textbf{Value of $\alpha$ (see section 4)} \\ \hline
		DBLP & 12,591 & 49,620 & 4 \\
		HepTh & 22,908 & 2,444,798 & 106 \\
		Cora & 23,166 & 89,157 & 4 \\
		HepPh & 28,093 & 3,148,447 & 112 \\
		CiteSeer & 384,413 & 1,736,145 & 4 \\
		Patents & 3,774,768 & 16,518,947 & 4 \\
		Gnutella1 & 36,682 & 88,328 & 2 \\
		Gnutella2 & 62,586 & 147,892 & 2 \\
		Facebook & 63,731 & 817,035 & 12 \\
		Prosperloans & 89,269 & 3,330,022 & 237 \\
		Amazon & 403,394 & 2,443,408 & 6 \\
		Pokec & 1,632,803 & 22,301,964 & 13 \\ \hline
		
	\end{tabular}
	\label{Tab1}
\end{table}

\begin{figure}[!t]
	\includegraphics[width=140mm, height=120mm]{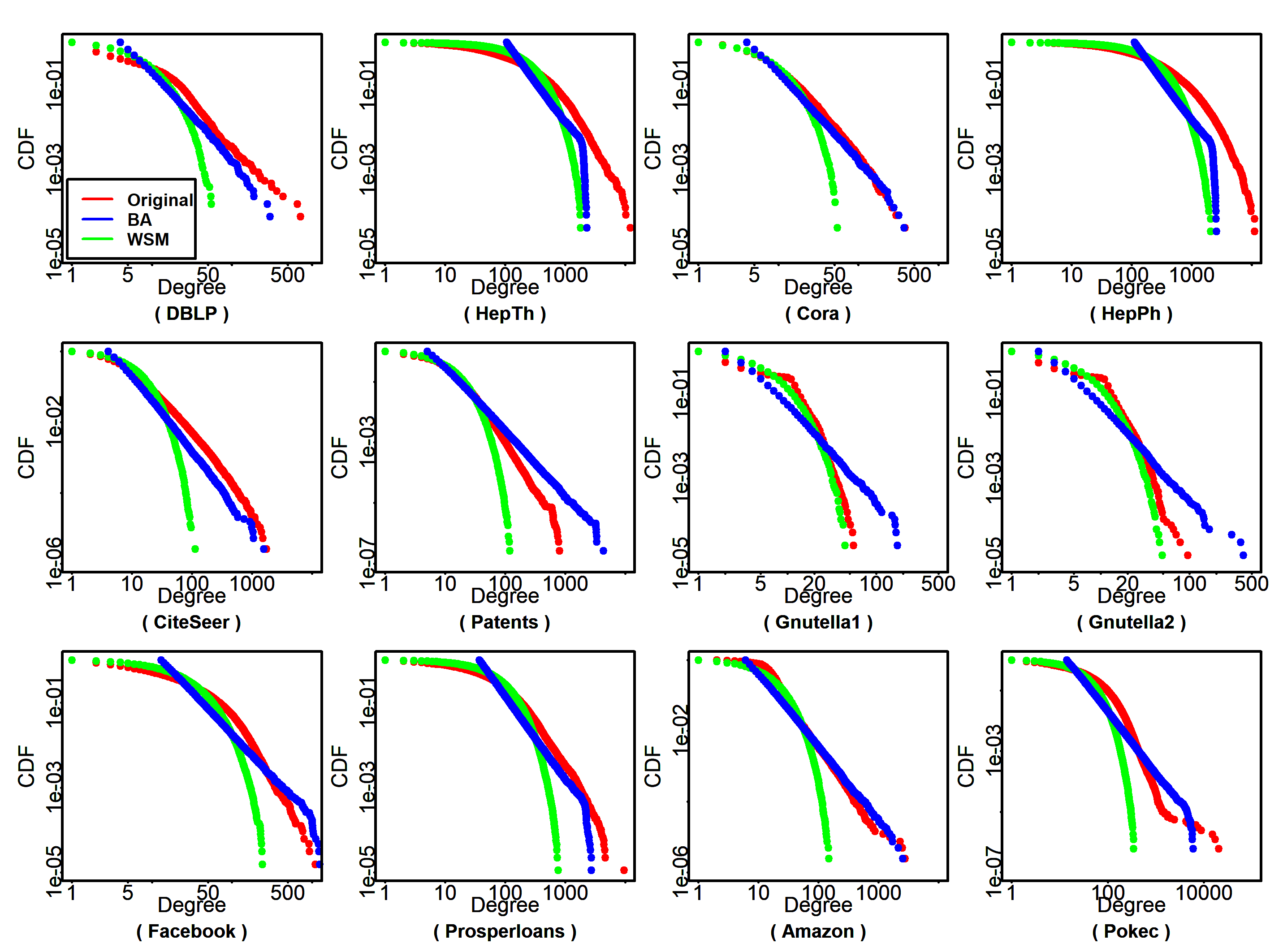}
	\centering
	\caption{Degree distribution of generated and original graphs. The value of $\alpha$ used for each dataset is given in the last column of Table 1.}
	\label{Result_Fig}
\end{figure}

We implement our model and generate networks corresponding to each real-network of Table~\ref{Tab1}. We set the value of $\Delta m_t=\alpha$ such that we have nearly the same number of edges in our network as that of in the original graph. It would be interesting to mention that if we apply the formula $m_t = (\alpha+1)(t-1)$ to find $\alpha$, where, $t$ is the number of nodes and $m_t$ is the number of edges, then the number of nodes and edges will be same for theoretical and practical networks. We generate networks and plot the Cumulative Distribution Function (CDF) of the degree of the original networks along with the networks generated by our model and Barrabasi-Albert model. The results are presented in Figure~\ref{Result_Fig}. First, these real-networks do not strictly follow the power-law degree distribution as clear from the initial part of the curve for small degree nodes. Second, we see that our model fits better than the BA model as we follow the distribution curve more than the BA model. It shows that we need to add some randomness in the graph for generating a weakly scale-free network. However, we believe that this randomness should be limited so that the degree distribution of the generated graph follows the original graph more precisely.

We apply the method presented in \cite{PL_Test} and give the values of power-law exponent and $k_{min}$ of original and generated graphs in Table~\ref{Tab2}. We also calculate the slope before $k_{min}$ for the original and generated graphs of WSM. We see that $\gamma >$ 3 for all the networks. It has already been observed in \cite{Cite1,Cite2} that $\gamma$ for citation networks fall in the range [3, 4], i.e., $\gamma \in $ [3,4] for citation networks. The table shows that $\gamma$ is almost equal to 3 for all the networks for the BA model whereas WSM can generate networks with varying values of $\gamma$.

\begin{figure}[t]
	\includegraphics[scale=0.55]{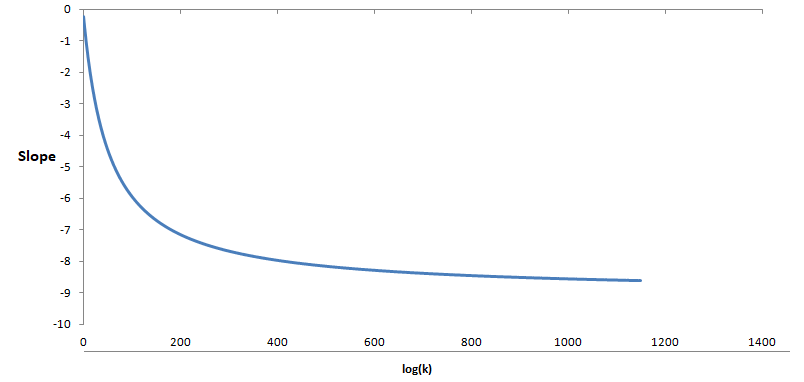}
	\centering
	\caption{Model generated slopes for $\alpha = 3$. Settling slope for ($\log(P_k) ~\text{ versus } \log(k)$) distribution is $-9$. }
	\label{fig:slopes}
\end{figure}

\begin{figure}[t]
	\includegraphics[scale=0.4]{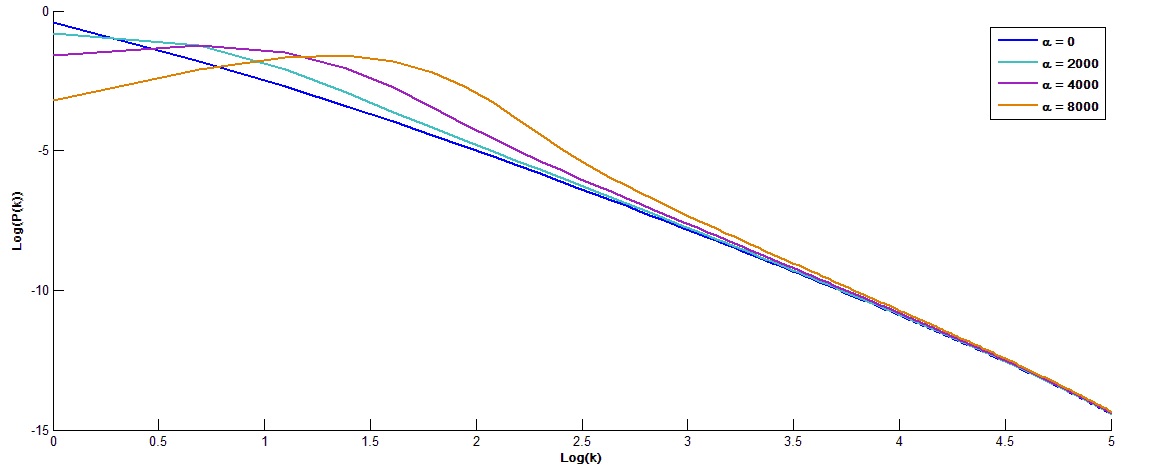}
	\centering
	\caption{Network is evolved using BA model first and then different number of edges edges added using the edge-step. Number of nodes used are $10000$. A portion of the graph is truncated. Graph marked by $0$, shows the distribution obtained by BA model. Graphs marked by $2000$, $4000$ and $8000$ are distributions obtained through adding different number of edges. Addition of edges flattens the first part of the degree distribution.($\log(P_k) ~\text{ versus } \log(k)$) distribution.}
	\label{fig:uniform}
\end{figure}

The degree distribution over logarithm scales ($\log(N_{k,t}) ~\text{ versus } \log(k)$) of chosen weakly scale-free networks starts with small slope (in magnitude) for low-degree nodes and becomes more and more sharp for high-degree nodes. After a particular degree value, say $k_{min}$, the distribution almost acquires a fixed sharp slope. Our model for a small choice of $\alpha$, generates weakly scale-free networks and the distribution given by the proposed model behaves similarly. The model gives degree distribution starting with a small slope, which becomes sharper, until at the $k_{min}(\alpha)$, it almost acquires a sharp fixed slope $-(2\alpha +3)$. Slope variation of the distribution is depicted in the Figure~\ref{fig:slopes}. In the beginning, the slope is small (in magnitude), which gradually becomes sharper and finally acquires the fixed value of $-9$. The BA model does not fit the first portion of the curve corresponding to low degrees. This initial section of the distribution is more uniform and cannot be obtained through preferential attachment alone. One possible solution to obtain this uniformity lies in blending randomness to preferential attachment, which can be obtained through the edge-step described in the model. When we apply the edge-step to the network which initially follows the power-law distribution, there is a high probability of nodes being chosen from groups marked by low degrees (see Figure~\ref{fig:uniform}). This is because each node has the same probability of selection and low-degree nodes are more in number. As a result, these chosen nodes will move into the next higher degree node group, increasing the rank of next higher-degree nodes and making the degree distribution more uniform. The portion of the curve which tends to have uniform distribution depends on the number $\alpha$.  When '$\alpha$' increases, $k_{min}$ shifts towards the right and the slope of portion after $k_{min}$ sharpens more, while the first portion characterized by flatness stretches rightwards. For generating these networks, the value of $\alpha$ is chosen as a small integer fixed for all time steps. Constant value gives the desired result and makes the formulation simple. A larger value makes a larger portion of the degree distribution uniform. So, choosing an appropriate value of $\alpha$ is desired.

\begin{table}
	\centering
	\caption{Values of power-law exponent $\gamma$ and $k_{min}$ of original and generated graphs.}
	\begin{tabular}{ |p{1.5cm}|p{0.9cm}|p{0.9cm}|p{0.9cm}|p{0.9cm}|p{0.9cm}|p{0.9cm}|p{0.8cm}|p{0.8cm}|}
		\hline
		\multirow{2}{*}{Dataset} &
		\multicolumn{3}{c|}{Original} &
		\multicolumn{3}{c|}{WSM} &
		\multicolumn{2}{c|}{BA}  \\ \cline{2-9}
		& \textbf{$\gamma$} & $k_{min}$ & Slope before $k_{min}$ & $\gamma$ & $k_{min}$ & Slope before $k_{min}$ & $\gamma$ & $k_{min}$ \\ \hline
		DBLP & 3.35 & 25 & 1.14 & 5.23 & 22 & 1.12 & 2.97 & 11 \\
		HepTh & 3.41 & 1244 & 0.97 & 5.46 & 895 & 0.77 & 2.95 & 109 \\
		Cora & 3.32 & 30 & 1.48 & 7.58 & 32 & 1.57 & 3.01 & 11 \\
		HepPh & 4.04 & 3226 & 0.91 & 5.48 & 953 & 0.72 & 2.96 & 111 \\
		CiteSeer & 3.05 & 74 & 1.95 & 7.47 & 51 & 1.75 & 3.07 & 19 \\
		Patents & 4.01 & 49 & 1.71 & 6.97 & 47 & 1.66 & 3.09 & 29 \\
		Gnutella1 & 4.92 & 11 & 1.44 & 5.89 & 18 & 1.39 & 3.32 & 15 \\
		Gnutella2 & 4.82 & 13 & 1.12 & 7.28 & 24 & 1.23 & 3.25 & 14 \\
		Facebook & 4.39 & 157 & 1.35 & 6.56 & 112 & 1.27 & 2.99 & 38 \\
		Pros.loans & 3.07 & 445 & 1.08 & 5.91 & 323 & 1.15 & 2.97 & 61 \\
		Amazon & 3.39 & 53 & 0.89 & 7.01 & 66 & 1.05 & 3.06 & 31 \\
		Pokec & 5.06 & 291 & 0.77 & 6.84 & 144 & 0.53 & 3.03 & 73 \\ \hline
		
	\end{tabular}
	\label{Tab2}
\end{table}

\subsection{Discussion}

We have implemented a two step approach to network evolution. The preferential attachment used in the node-step tends to make the nodes distribution follow the power-law. The edge creation probability function gives uniform probability to each node, which tends the network to acquire uniform degree distribution, especially the portion for low-degree nodes. In the case when $\Delta m_t$ varies during each time step $t$, the edge-step dominates the node-step and the distribution becomes uniform, implying that the network is no longer a scale-free network, rather, is a random network. We see in Figure~\ref{Result_Fig} that our model closely follows the original networks, better than the BA model but not perfectly. We argue that our model fits the overall trend of the degree distribution, however, it needs tweaking for matching the fine details of a distribution. In the future, we would like to invest in this direction and also consider other metrics for comparing our model with the real networks.

\section{Related Work}
There have been many studies \cite{PL1, PL2, PL3, PL4, PL5, PL6, PL7, PL8, PL9} to explore the structure of complex online networks. For example, the authors in \cite{PL6, PL7} study the properties of real networks by analyzing large scale Online Social Networks (OSNs) and discover link symmetries, scale-free degree distributions, clustering phenomena, and community formations. Golder et al. \cite{PL8} analyze the Facebook network by studying the messaging pattern between friends and report that the degree distribution of the Facebook network follows a power-law. Kumar et al. \cite{PL9} study the structure and dynamics of Online Social Networks over time and proposes a generative model for OSNs. However, a very recent study \cite{SF_rare} analyzes many real-world networks and concludes that scale-free networks are rare and only a small number of real-world graphs follow a power-law degree distribution. The study also finds that a majority of real networks are weakly scale-free. In \cite{D4} authors presented a method of growing directed networks through mixing preferential and random linking. Preferential linking is used to connect a new node to the network. Then, the edges emerge in the network and nodes for these edges are chosen through the preferential attachment method. In the last step,  the edges are created with end-nodes selected at random.

Generative models for complex networks produce graphs that typically satisfy the scale-free degree distribution. For example, the preferential attachment model \cite{Barabasi}, one of the best-known models for scale-free networks, simulates this property. Real-world modeling instances motivated the proposal of generalizations of the preferential attachment model, e.g., \cite{PA1,PA2, PA3, PA4, PA5, PA6, PA7, PA8, PA9, PA10, PA11, PA12}. A common characteristic of many of these models is the presence of the same attachment rule for all the nodes of the network. However, this hypothesis is not always realistic  and the graphs generated by these models do not follow a weakly scale-free degree distribution. In \cite{PA10_Win1}, Magner et al. introduced a model in which new vertices choose the nodes for their links within time-based windows. Inspired by the idea of introducing windows, the authors in \cite{PA11_Win2} formulate their idea. They apply the preferential attachment rule to any node but re-enforce this rule with a uniform choice for the most recent nodes added to the network.

\section{Conclusion}
In this paper, we propose a generative model for generating weakly scale-free networks. The model generated degree distribution is similar in trend with real network data for the whole range of degree values $k>0$. We provide a comprehensive mathematical analysis of the model in the discrete domain. We compare the model with real networks and find that our graphs closely match with the original graphs in their degree distribution.

\nocite{*}
\bibliographystyle{abbrvnat}
% use the following instead if you encounter problems 
%\bibliographystyle{alpha}
\bibliography{sample-dmtcs}
\label{sec:biblio}

\appendices
\section{General form of Recurrence Relation for degree distribution}
\label{Appendix A}
\setcounter{equation}{0}
\renewcommand{\theequation}{A.\arabic{equation}}

A general form of the recurrence relation for degree distribution will be derived here. Recurrence formula writes $P_{k,t+1}$, rank of nodes of degree $k$ at time $t+1$, in terms of $P_{h,t}$ for $h=k,k-1, \dots $. Network evolution is based on both the node- and edge- steps. The formulation is valid whether $\Delta m_t$ is constant for all time steps or varies by the rule (given in \ref{Appendix B}).

\textbf{The relation for $k =1 $.} \newline

Using Equation  (\ref{Edgek1}) repeatedly for $l=1,2,\dots, \Delta m_t$, we get
\[
\begin{split}
N_{k,t+1} = \bigl(1-\frac{2}{t+1}\bigr)^{\Delta m_t}N_{k,t+1,0}.
\end{split}
\]

Now using the equation (\ref{Nodek1}), $N_{k,t+1}$ after the execution of the node and edge-step:
\[
N_{k,t+1} = \bigl(1-\frac{2}{t+1}\bigr)^{\Delta m_t}\bigl[\bigl(1-\frac{k}{2m_t}\bigr)N_{k,t} + 1 \bigr],
\]

and writing the equation in terms of ranks $P_{k,t}$'s, following form is obtained
\begin{equation}\label{Genk1}
(t+1)P_{k,t+1} = \bigl(1-\frac{2}{t+1}\bigr)^{\Delta m_t}\bigl[tP_{k,t} + 1 - \frac{k}{2m_t}tP_{k,t} \bigr].
\end{equation}

\textbf{The derivation of recurrence relation for $ k > 1$.}\newline
We will write general form of the recurrence relation for $k > 1$.
Let $\Phi = \frac{2}{t+1}$ and $\Psi=1 - \frac{2}{t+1}$.

Then rewriting the recurrence relation (\ref{EdgekM}), for for $l=1$ and  general $l$, in the following form:
\[
\begin{split}
&N_{k,t+1,1} = \Psi N_{k,t+1,0} + \Phi N_{k-1,t+1,0},\\
&N_{k,t+1,l} = \Psi N_{k,t+1,l-1} + \Phi N_{k-1,t+1,l-1}.
\end{split}
\]

Next we will write $N_{k,t+1,l}$ in terms of $N_{h,t+1,0}$ for $h=k,k-1, \dots, k-l$
\[
\begin{split}
&N_{k,t+1,2} = \Psi N_{k,t+1,1} + \Phi N_{k-1,t+1,1} \text{, for } k  \geq 2,\\
&N_{k,t+1,2} = \Psi^2 N_{k,t+1,0} + 2\Psi \Phi N_{k-1,t+1,0} + \Phi^2 N_{k-2,t+1,0} \text{, for } k  \geq 3,\\
&N_{k,t+1,3} = \Psi^3 N_{k,t+1,0} + 3\Psi^2 \Phi N_{k-1,t+1,0} + 3\Psi \Phi^2 N_{k-2,t+1,0}+ \Phi^3 N_{k-3,t+1,0} \text{, for } k  \geq 4.
\end{split}
\]

Define $q=\min(k-1,\Delta m_t)$ and write $N_{k,t+1}$ in a brief form,

\[
N_{k,t+1} = \sum_{n=0}^{q} C_{k-n,\Delta m_t}^{'}N_{k-n,t+1,0}.
\]

Now, using the relation (\ref{Nodek})
\[
N_{k,t+1,0} =\bigl(1-\frac{k}{2m_t} \bigr) N_{k,t} + N_{k-1,t}\frac{k-1}{2m_t},
\]

we get $N_{k,t+1}$ in terms of $N_{k-n,t}$'s which represent network configuration at previous sampling time.
\[
\begin{split}
N_{k,t+1} = \sum_{n=0}^{q} C_{k-n,\Delta m_t}^{'} \bigl(1-\frac{k-n}{2m_t} \bigr) N_{k-n,t} + \sum_{n=0}^{q-1} C_{k-n,\Delta m_t}^{'}  \frac{k-n-1}{2m_t}N_{k-n-1,t} \\
+ \bigl[C_{k-q,\Delta m_t}^{'}  \frac{k-q-1}{2m_t}N_{k-q-1,t}\bigr].
\end{split}
\]
Last term in square brackets occurs only when $q =\Delta m_t <   k-1$. During the application of the limit, $\lim t \to \infty$, this situation can occur only when $\Delta m_t$ is a fixed number. In the other case, where $\Delta m_t$ varies, $\Delta m_t \to \infty$, this term does not appear in the recurrence relation.
Simplifying further,we get
\[
\begin{split}
N_{k,t+1} &= C_{k,\Delta m_t}^{'}\bigl(1-\frac{k-n}{2m_t} \bigr) N_{k,t} + \sum_{n=1}^{q} C_{k-n,\Delta m_t}^{'} \bigl(1-\frac{k-n}{2m_t} \bigr) N_{k-n,t},   \\
&=\sum_{n=0}^{q-1} C_{k-n,\Delta m_t}^{'}\frac{k-n-1}{2m_t}N_{k-n-1,t}  + C_{k-q,\Delta m_t}^{'}\frac{k-q-1}{2m_t}N_{k-q-1,t}.
\end{split}
\]

\[
\begin{split}
\implies  N_{k,t+1} = \bigl( 1 - \frac{k}{2m_t} \bigr)C_{k,\Delta m_t}^{'}N_{k,t}+ \sum_{n=1}^{q} \bigl[\bigl( 1 - \frac{k-n}{2m_t}\bigr)C_{k-n,\Delta m_t}^{'} + \frac{k-n}{2m_t}C_{k-n+1,\Delta m_t}^{'}\bigr]N_{k-n,t}\\
+\bigl[\frac{k-\Delta m_t-1}{2m_t}C_{k-\Delta m_t,\Delta m_t}^{'}N_{k-\Delta m_t-1} \bigr].
\end{split}
\]

Using the relation $P_{k,t} = \frac{N_{k,t}}{t}$,we get
\[
\begin{split}
(t+1)P_{k,t+1} = \bigl( 1 - \frac{k}{2m_t} \bigr)C_{k,\Delta m_t}^{'}tP_{k,t}+ \sum_{n=1}^{q} \bigl[\bigl( 1 - \frac{k-n}{2m_t}\bigr)C_{k-n,\Delta m_t}^{'} + \frac{k-n}{2m_t}C_{k-n+1,\Delta m_t}^{'}\bigr]tP_{k-n,t}\\
+\frac{k-\Delta m_t-1}{2m_t}C_{k-\Delta m_t,\Delta m_t}^{'}tP_{k-\Delta m_t-1}.
\end{split}
\]

In the brief form,
\begin{equation}\label{Genk}
(t+1)P_{k,t+1} = \sum_{n=0}^{q} tC_{k-n,\Delta m_t}P_{k-n,t},
\end{equation}

where coefficients $C_{k-n,\Delta m_t}$ are defined as \newline
for $n=0$, we get
\[
C_{k,\Delta m_t} =  \bigl( 1 - \frac{k}{2m_t} \bigr)C_{k,\Delta m_t}^{'}.
\]

For $1<n\leq q$, expression becomes as:
\[
C_{k-n,\Delta m_t} = \bigl( 1 - \frac{k-n}{2m_t} \bigr)  C_{k-n,\Delta m_t}^{'} + \frac{k-n}{2m_t} C_{k-n+1,\Delta m_t}^{'},
\]

and if $ \Delta m_t < k-1$, we get

\[
C_{k-\Delta m_t,\Delta m_t} =  \frac{k-\Delta m_t-1}{2m_t}C_{k-\Delta m_t,\Delta m_t}^{'}.
\]

Next we derive formulas for $C_{k-n,l}^{'}$'s and $C_{k-n,l}$'s. \newline
\textbf{Key observations:}
Before deriving exact form of these coefficients, we list few key points about these coefficients.
\[
C_{k,l}^{'} = \Psi C_{k ,l-1}^{'}  \text{ for } l = 1,2,\dots.
\]

It can be seen that $C_{k-n,l}^{'} \text{ for } n = 1,2,\dots $ term appears first time in the expression after the addition of  $l = n$ edges. For $l < n,  ~ ~ C_{k-n,l}=0$.
Putting in the mathematical form, for $n=1,2...$ and $ n < k$,
\[
\begin{split}
&C_{k-n,l}^{'} = 0, \text{ for } l < n,\\
&C_{k-n,n}^{'} = \Phi ^n. \\
\end{split}
\]

Recurrence relation for coefficients is
\begin{equation} \label{CoeefRecurr}
C_{k-n,l}^{'} = \Psi C_{k-n ,l-1}^{'} + \Phi C_{k-n+1 ,l-1}^{'} \text{ for } l > n.
\end{equation}

Now we derive exact form of each coefficient $C_{k-n,\Delta m_t}\text{ for } n = 0,1,2,\dots$ \newline
$\bm{ C_{ k, \Delta m_t}}$
\[
\begin{split}
C_{k,l}^{'} = \Psi^l  \text { for } l=1,2,...,\Delta m_t,\\
\end{split}
\]

\begin{equation}\label{Coeef1}
\implies C_{k,\Delta m_t} = \bigl( 1 - \frac{k}{2m_t} \bigr) \bigl(1 - \frac{2}{t+1}\bigr)^{\Delta m_t}.
\end{equation}

$\bm{ C_{ k-1, \Delta m_t} \text{ for }  k > 1}$

\[
\begin{split}
C_{k-1,\Delta m_t} = \bigl( 1 - \frac{k-1}{2m_t} \bigr)  C_{k-1,\Delta m_t}^{'} + \frac{k-1}{2m_t} C_{k,\Delta m_t}^{'}. \\
\end{split}
\]

First we find an expression for $C_{k-1,\Delta m_t}^{'}$.
\[
\begin{split}
&C_{k-1,l}^{'} = b_l^{k-1} \Psi^{l-1} \Phi \text{ with } b_l^{k-1} = l, \text { for } l=1,2,...,\Delta m_t, \text{ using the relation \ref{CoeefRecurr}}\\
\implies &C_{k-1,\Delta m_t}^{'} = \Delta m_t \Psi^{\Delta m_t-1} \Phi, \\
\implies &C_{k-1,\Delta m_t} = \bigl( 1 - \frac{k-1}{2m_t} \bigr)  \Delta m_t \Psi^{\Delta m_t-1} \Phi  + \frac{k-1}{2m_t} \Psi^{\Delta m_t}. \\
\end{split}
\]
Finally,
\begin{equation}\label{Coeef2}
C_{k-1,\Delta m_t} = \bigl( 1 - \frac{k-1}{2m_t} \bigr)  \Delta m_t \bigl(1 - \frac{2}{t+1}\bigr)^{\Delta m_t-1} \frac{2}{t+1}  + \frac{k-1}{2m_t} \bigl(1 - \frac{2}{t+1}\bigr)^{\Delta m_t}. \\
\end{equation}

$\bm{ C_{ k-2, \Delta m_t}\text{ for }  k > 2}$ \newline

\[
C_{k-2,\Delta m_t} = \bigl( 1 - \frac{k-2}{2m_t} \bigr)  C_{k-2,\Delta m_t}^{'} + \frac{k-2}{2m_t} C_{k-1,\Delta m_t}^{'}, \\
\]

\[
\begin{split}
&C_{k-2,1}^{'} = 0 ,
C_{k-2,l}^{'} = b_l^{k-2} \Psi^{l-2} \Phi^2  \text{ for }  l=2,3,....\Delta m_t,\\
\text{with }  &b_2^{k-2}= 1, ~~ 	b_{l+1}^{k-2} = b_{l}^{k-2} + b_l^{k-1} \text{ for } l=2,3,\dots,\text{ using the relation \ref{CoeefRecurr}}. \\\\
\implies  &b_{l+1}^{k-2} = b_{l}^{k-2} + l \text{ for } l=2,3,\dots\\
\implies  &b_{l}^{k-2} = \frac{l(l-1)}{2},\\
\text{Now } &C_{k-2,l}^{'} = \frac{l(l-1)}{2} \Psi^{l-2} \Phi^2  \text{ for }  l=2,3,\dots\Delta m_t,\\
\implies  &C_{k-2,\Delta m_t}^{'} = \frac{\Delta m_t(\Delta m_t-1)}{2} \Psi^{\Delta m_t-2} \Phi^2.
\end{split}
\]
Finally we get,
\begin{equation}\label{Coeef3}
C_{k-2,\Delta m_t} = \bigl( 1 - \frac{k-2}{2m_t} \bigr)  C_{k-2,\Delta m_t}^{'} + \frac{k-2}{2m_t} C_{k-1,\Delta m_t}^{'}. \\
\end{equation}

$\bm{ C_{ k-n, \Delta m_t} \text{ for }  k > n  ,\text{ for } n=3,4...}$
\\

\begin{equation}\label{Coeefn}
C_{k-n,\Delta m_t}^{'} = b_{\Delta m_t}^{k-n} \Psi^{\Delta m_t-n} \Phi^n.
\end{equation}

Determination of orders of these terms is sufficient. $b_{\Delta m_t}^{k-n}$ is of  $\mathcal{O}( {\Delta m_t}^n)$. Order of $C_{ k-n, \Delta m_t}$ and $C_{k-n,\Delta m_t}^{'}$ are the same. Order of $C_{ k-n, \Delta m_t}$ is $\mathcal{O}( {\Delta m_t}^n) \times \mathcal{O}(\frac{1}{t^n}) = \mathcal{O} ( ({\frac{\Delta m_t}{t}}) ^n)$.

\subsection{$\Delta m_t = \alpha$}
This section derives recurrence relation for ranks $P_k$'s when number of edges is a fixed constant $\alpha$ for each time step $t$.\newline
\textbf{$k=1$}

Consider eqn. (\ref{Genk1}) for $\Delta m_t=\alpha$

\[
(t+1)P_{k,t+1} = \bigl(1-\frac{2}{t+1}\bigr)^{\alpha}\bigl[\bigl(1-\frac{k}{2(\alpha+1)(t-1)}\bigr)tP_{k,t} + 1 \bigr]
\]

\[
\begin{split}
\implies &tP_{k,t+1} + P_{k,t+1} = \bigl(1-\frac{2}{t+1}\bigr)^{\alpha}\bigl[ tP_{k,t} - \frac{kt}{2(\alpha+1)(t-1)}P_{k,t} + 1 \bigr],\\
\implies &tP_{k,t+1} + P_{k,t+1} = \bigl(1-\frac{2\alpha}{t+1}+ \dots \bigr)\bigl[ tP_{k,t} - \frac{kt}{2(\alpha+1)(t-1)}P_{k,t} + 1 \bigr],\\
\implies &tP_{k,t+1} + P_{k,t+1} = tP_{k,t} - \frac{kt}{2(\alpha+1)(t-1)}P_{k,t} + 1 - \frac{2\alpha}{t+1}tP_{k,t} +\dots.
\end{split}
\]

On applying limit $\lim t \to \infty$ this equation acquires the form
\[
P_{k} =-\frac{k}{2(\alpha+1)}P_{k} + 1 - 2\alpha P_{k},
\]

or
\[
P_{k} =  1- \bigl[2\alpha  + \frac{k}{2(\alpha+1)} \bigr]P_{k}.
\]

\begin{equation}%\label{UEdgeEqnM4}
\implies P_{1} =  \frac{2(\alpha+1)}{ 4\alpha^2 + 6\alpha +3}.
\end{equation}

\textbf{$k > 1$} \newline

We will find expressions for coefficients $C_{ k-n, \Delta m_t} ~ n = 0,1,2 \dots  \text{ for } \Delta m_t = \alpha$ and evaluate the recurrence relation eqn. (\ref{Genk}).\newline

$\bm{ C_{ k, \alpha}}$
\[
\begin{split}
&C_{k,\alpha} =  \bigl(1 - \frac{2}{t+1}\bigr)^{\alpha} \bigl( 1 - \frac{k}{2m_t} \bigr),\\
\implies &tC_{k,\alpha} = \bigl(1 - \frac{2}{t+1}\bigr)^{\alpha} \bigl( t - k\frac{t}{2(\alpha+1)(t-1)} \bigr), \\
\implies  &tC_{k,\alpha} = \bigl(1 - \frac{2\alpha}{t+1}+ \dots\bigr) \bigl( t - k\frac{t}{2(\alpha+1)(t-1)} \bigr). \\
\implies &tC_{k,\alpha}P_{k,t} =   tP_{k,t} - k\frac{t}{2(\alpha+1)(t-1)}P_{k,t}   - \frac{2\alpha t}{t+1}P_{k,t}+ \dots.
\end{split}
\]
$tP_{k,t}$ will be cancelled out by the term on L.H.S.(\ref{Genk}).  $tC_{k,\alpha}P_{k,t}$ approaches to
$- \bigl(\frac{k}{2(\alpha+1)}+2\alpha \bigr)P_{k}$.\newline

$\bm{ C_{ k-1, \alpha}} $ \newline

\[
\begin{split}
&C_{k-1,\alpha} = \alpha \bigl( 1 - \frac{k-1}{2m_t} \bigr)   \bigl(1 - \frac{2}{t+1}\bigr)^{\alpha-1} \frac{2}{t+1}  + \frac{k-1}{2m_t} \bigl(1 - \frac{2}{t+1}\bigr)^{\alpha}, \\
\implies &tC_{k-1,\alpha} = \alpha \bigl( 1 - \frac{k-1}{2m_t} \bigr)   \bigl(1 - \frac{2}{t+1}\bigr)^{\alpha-1} \frac{2t}{t+1}  + (k-1)\frac{t}{2(\alpha+1)(t-1)} \bigl(1 - \frac{2}{t+1}\bigr)^{\alpha} \\
\end{split}
\]

Term $tC_{k-1,\alpha} P_{k-1,t}$ approaches to $(2\alpha +\frac{k-1}{2(\alpha+1)})P_{k-1}$. \newline

$\bm{ C_{ k-2, \alpha}\text{ for } k > 2}$
\newline

\[
\begin{split}
&C_{k-2,\alpha}^{'} = \frac{\alpha(\alpha-1)}{2} \Psi^{\alpha-2} \Phi^2,  \\
\implies &C_{k-2,\alpha}^{'} = \frac{\alpha(\alpha-1)}{2} \bigl (1-\frac{2}{t+1} \bigr)^{\alpha-2} \bigl(\frac{2}{t+1}\bigr)^2,  \\
\text{with already determined }	&C_{k-1,\alpha}^{'} = a_l^{k-1} \Psi^{\alpha-1} \Phi, \\
\implies &tC_{k-2,\alpha} = \bigl( 1 - \frac{k-2}{2(\alpha+1)(t-1)} \bigr)  tC_{k-2,\alpha}^{'} +  \frac{k-2}{2(\alpha+1)(t-1)} tC_{k-1,\alpha}^{'}. \\
\end{split}
\]
Maximum order term in the $tC_{k-2,\alpha}$ is of order $\mathcal{O}(\frac{1}{t})$ .The term $tC_{k-2,\alpha}$ vanishes to $0$ on taking limit $\lim_{t \to \infty}$. \newline

$\bm{ C_{ k-n, \alpha}} \text{ for } k > n$  for $n=3,4,...$ \newline
$tC_{ k-n, \alpha}$ is of order $\frac{1}{t^{n-1}}$. $tC_{ k-n, \alpha}$ for $n=3,4,...$ vanishes on applying the limit.

Now we have all limits $\lim_{t \to \infty}tC_{k-n,\alpha}$ required to determine limiting value of the equation (\ref{Genk}). Equation  (\ref{Genk}) on applying limit becomes

\[
P_{k} =  \bigl[ 2\alpha+ \frac{k-1}{2(\alpha+1)} \bigr]P_{k-1}- \bigl[2\alpha  + \frac{k}{2(\alpha+1)} \bigr]P_{k},
\]

\begin{equation}%\label{UEdgeEqnM2}
\implies P_{k} = \frac{k+ 4\alpha^2 + 4\alpha -1}{k + 4\alpha^2 + 6\alpha +2}P_{k-1}.
\end{equation}

\section{Variable $\Delta m_t $}
\label{Appendix B}
\setcounter{equation}{0}
\renewcommand{\theequation}{B.\arabic{equation}}

Here, for the sake of completion, we will discuss the case when $\Delta m_t$ varies with time $t$ and find recurrence relation. It varies at each time step $t$, with $\Delta m_t+1$ is proportional to the fraction of the total number of edges by the number of nodes. Recurrence relation will be derived only formally.
\par

\subsection{The number of edges in edge-step at time step $t$}

The number of edges at the completion of time step $t$ (both node and edge-step completed ) is $m_t$. $\Delta m_t$ is the number of edges to be added during the edge-step. As mentioned earlier, this number is determined by the fraction of edges by number of nodes. $\Delta m_t =  \lceil \frac{\beta m_t }{t}\rceil - 1 $, with parameter $\beta$ as a real number greater than $1$. For the current analysis, $\beta \in(1,2) $.
After the node-step, total number of number of edges are $m_{t + 1}$.  $m_{t+1}$, the total number of edges after the completion of both node- and edge- steps, is given by the relation

\[
m_{t+1} = m_t + \Delta m_t + 1,
\]
We select $\beta$ such that $\frac {\beta m_t}{t} >> 1$, $\Delta m_t \approx \frac {\beta m_t}{t}$. \\
It implies that,
\[
\begin{split}
&m_{t+1} \approx  m_t (1 + \frac{ \beta}{t})\\
\implies &m_{t+1} \approx m_1 (1 + \frac{ \beta}{t}) (1 + \frac{ \beta}{t-1})... (1 + \frac{ \beta}{1}),\\
\implies &m_{t+1} \approx m_1  \frac{(\beta + t) (\beta + t -1 )(\beta + t -2 )... (\beta +1 )}{t!},\\
\implies &m_{t+1} \approx m_1  \frac{\Gamma(\beta + t+1)}{\Gamma(t+1)\Gamma(\beta+1)}.
\end{split}
\]

Stirling's formula,
\[
\Gamma(t+1) \approx \sqrt{2 \pi t}  \bigl(\frac{t}{e}\bigr)^t
\]
which gives good approximation for the factorials of large number, can be used here to get simplified form of the above expression.
Using this approximation, expression for  $m_{t+1}$ reduces to
\[
m_{t+1} \approx \frac{1}{\sqrt{2 \pi \beta ^ {3\beta}}} \frac{(\beta + t)^{\beta + t + \frac{1}{2}}}{t ^{t + \frac{1}{2}}},
\]

or
\[
m_{t+1} \approx \frac{\bigl({1+\frac{\beta}{t}}\bigr)^{\beta + t + \frac{1}{2}}}{\sqrt{2 \pi \beta ^ {3\beta}}} t^{\beta},
\]

\begin{equation}\label{mtf1}
m_{t+1} \approx  m^*_1 t^{\beta}.
\end{equation}

where $ m^*_1  =  \frac{m_1 e^{\beta}}{\sqrt{2 \pi \beta ^ {3\beta}}}$.

%Thus for large $t$s.
Thus,
\begin{equation}\label{mtf2}
\lim_{t \to \infty} m_{t} =  \lim_{t \to \infty} m^*_1  t^{\beta}.
\end{equation}

This formulation is valid only for $\Delta m_t >> 1$. It implies that $\beta > 1$.

\subsection{Recurrence relation for degree distribution}
\textbf{$k =1 $}\newline

Consider the equation \ref{Genk1} once again.
\[
(t+1)P_{k,t+1} = \bigl(1-\frac{2}{t+1}\bigr)^{\Delta m_t}\bigl[tP_{k,t} + 1 - \frac{k}{2m_t}tP_{k,t} \bigr].
\]

Formally speaking, if equation is divided by $t+1$ and limit $\lim_{t \to \infty}$ is applied, we get $P_{k,t+1} \to P_{k,t}$. Let us write equation into the form where we can cancel the highest order terms.

\[
\bigl[(t+1)  - \bigl(1-\frac{2}{t+1}\bigr)^{\Delta m_t} \bigl(1-\frac{k}{2m_t}\bigr)t \bigr]P_{k,t+1} = \bigl(1-\frac{2}{t+1}\bigr)^{\Delta m_t} \bigl[ 1 - \frac{k}{2m_t}tP_{k,t} \bigr].
\]

The $\lim_{t \to \infty} \bigl(1-\frac{2}{t+1}\bigr)^{\Delta m_t} = \lim_{t \to \infty} \exp({\Delta m_t \ln{\bigl(1-\frac{2}{t+1}\bigr)}})=1$ for $\beta < 2$. On applying the limit, R.H.S. becomes $1$. Let us now evaluate L.H.S. limit.

\[
\begin{split}
&\lim_{t \to \infty} \bigl[(t+1)  - \bigl(1-\frac{2}{t+1}\bigr)^{\Delta m_t} \bigl(1-\frac{k}{2m_t}\bigr)t \bigr] \\
&=\lim_{t \to \infty} \bigl[(t+1)  - \bigl(1-\frac{2}{t+1}\bigr)^{\Delta m_t} t \bigr]\\
&=\lim_{t \to \infty} \bigl[(t+1)  - \bigl(1-\frac{2\Delta m_t }{t+1}+ \dots \bigr) t \bigr]\\
&=\lim_{t \to \infty} \bigl[1  +   2\Delta m_t +\dots \bigr]\\
\end{split}
\]
is the reduced form of the L.H.S. limit.

Equating both limits,
\[
\implies	\lim_{t \to \infty} \bigl[1  +  2 \Delta m_t+\dots \bigr]P_{k,t+1} =1.
\]

It can be seen that for $\beta \in(1,2) $, the highest order term in the expression $\bigl[(t+1)  - \bigl(1-\frac{2}{t+1}\bigr)^{\Delta m_t} \bigl(1-\frac{k}{2m_t}\bigr)t \bigr]$ is $t^{\beta-1}$, whereas the R.H.S. of the equation has order $1$.

This implies that

\[
P_{1} \to 0.
\]

Now consider the \textbf{case for $k > 1$} . We will evaluate limiting values of coefficients $tC_{k-n,\Delta m_t}, n= 0,1,\dots$ \newline

$\bm{ C_{ k, \Delta m_t}}$\newline
\[
\begin{split}
C_{k,\Delta m_t} = \bigl( 1 - \frac{k}{2m_t} \bigr) \bigl(1 - \frac{2}{t+1}\bigr)^{\Delta m_t}.\\
\end{split}
\]

Gathering terms with $k$ on L.H.S, coefficient of $P_{k,t}$ in (\ref{Genk}) for very large $t$ becomes $((t+1) - tC_{k,\Delta m_t})$, with
\[
\begin{split}
(t+1) - tC_{k,\Delta m_t}&= \bigl[(t+1)  - \bigl(1-\frac{2}{t+1}\bigr)^{\Delta m_t} \bigl(1-\frac{k}{2m_t}\bigr)t \bigr] \\
&=\bigl[(t+1)  - \bigl(1-\frac{2}{t+1}\bigr)^{\Delta m_t} t \bigr]\\
&=\bigl[(t+1)  - \bigl(1-\frac{2\Delta m_t }{t+1}+ \dots \bigr) t \bigr]\\
&=\bigl[1  +  \frac{2t \Delta m_t}{t+1}+\dots \bigr]\\
\end{split}
\]

Limiting value of the $[(t+1) - tC_{k,\Delta m_t}]P_{k,t}$ becomes:
\[
\lim_{t \to \infty} \bigl[1  +  \frac{2t \Delta m_t}{t+1}+\dots \bigr]P_{k,t} = \lim_{t \to \infty} \bigl[1  +  {2 \Delta m_t}+\dots \bigr]P_{k}.
\]

The highest order term in the expression $\bigl[(t+1)  - \bigl(1-\frac{2}{t+1}\bigr)^{\Delta m_t} \bigl(1-\frac{k}{2m_t}\bigr)t \bigr]$ is of order $\mathcal{O}{(\Delta m_t)}$ or of order $t^{\beta - 1}$. Thus the coefficient of is of $P_{k,t}$ has order $\mathcal{O}{(\Delta m_t)}$.\newline

$\bm{ C_{ k-1, \Delta m_t}}$
\newline

This coefficient exists for $ k > 1$. \newline

\[
\begin{split}
C_{k-1,\Delta m_t} = \bigl( 1 - \frac{k-1}{2m_t} \bigr)  \Delta m_t \bigl(1 - \frac{2}{t+1}\bigr)^{\Delta m_t-1} \frac{2}{t+1}  + \frac{k-1}{2m_t} \bigl(1 - \frac{2}{t+1}\bigr)^{\Delta m_t}. \\
\end{split}
\]

Consider $tC_{k-1,\Delta m_t}$,
\[
\begin{split}
tC_{k-1,\Delta m_t} = \bigl( 1 - \frac{k-1}{2m_t} \bigr)  \Delta m_t \bigl(1 - \frac{2}{t+1}\bigr)^{\Delta m_t-1} \frac{2t}{t+1}  + \frac{(k-1)t}{2m_t} \bigl(1 - \frac{2}{t+1}\bigr)^{\Delta m_t}. \\
\end{split}
\]

As already shown that $\bigl(1 - \frac{2}{t+1}\bigr)^{\Delta m_t} \to 1$, second term in the sum will approach to $0$. The first term contains $2\Delta m_t$ as a maximum order term. Thus $2\Delta m_t$ is maximum order coefficient of $P_{k-1,t}$. In short, $O(t^{\beta-1})$ is the order of $tC_{k-1,\Delta m_t}P_{k-1,t}$.\newline

$\bm{ C_{ k-2, \Delta m_t}\text{ for } k > 2}$\newline

This coefficient exists for $ k > 2$.

\[
C_{k-2,\Delta m_t} = \bigl( 1 - \frac{k-2}{2m_t} \bigr)  C_{k-2,\Delta m_t}^{'} + \frac{k-2}{2m_t} C_{k-1,\Delta m_t}^{'}. \\
\]

Consider $tC_{k-2,\Delta m_t}$.
Highest order term in the expression is of $O( \frac{\Delta m_t(\Delta m_t-1)}{2} t \Phi^2) = \mathcal{O}(t^{2\beta-3})$. Thus maximum order coefficient of $P_{k-2,t}$ has the order $\mathcal{O}(t^{2\beta-3})$. This term when divided by the term of the order $\mathcal{O}(t^{\beta-1})$, results into a term of order $t^{\beta-2}$,  which will approach to $0$ for $\beta <2$.\newline

$\bm{ C_{ k-n, \Delta m_t}\text{ for } k > n}$ for $n=3,4,...$\newline
\\
$C_{k-n,\Delta m_t} = a_{\Delta m_t}^{k-n} \Psi^{\Delta m_t-n} \Phi^n$  with $a_{\Delta m_t}^{k-n}$ is of  $\mathcal{O}( {\Delta m_t}^n)$.\newline

It is sufficient to find only orders of these terms. Order of $tC_{ k-n, \Delta m_t}$ is  $\mathcal{O}(t^{n(\beta-1)-1})$. On dividing the equation by $t^{\beta -1}$, we get coefficients of $P_{ k-n,t}$ of order $t^{(n-1)(\beta-2)}$  that vanishes to $0$ for $ \beta < 2$ on applying the limit $\lim_{t \to \infty}$. Thus, on applying
$\lim_{t \to \infty}$ to (\ref{Genk}), following expression is obtained:

\[
\lim_{t \to \infty} \bigl[1  +  2 \Delta m_t+\dots \bigr]P_{k} = \lim_{t \to \infty}[ 2\Delta m_t]P_{k-1}.
\]

\[
P_{k} =  \lim_{t \to \infty}[ \frac{2 \Delta m_t}{2 \Delta m_t  + 1} ]P_{k-1}.
\]

Thus for $k = 2,3,\dots$
\[
P_{k} =  P_{k-1}.
\]
or
\[
P_{k} =  P_{1}.
\]

At each time step $P_{k,t}$ must satisfy the constraint $\sum_{k=1}^{\infty} P_{k,t} = 1$. For a moment, assume that $P_{k}$ is a stationary degree distribution. Then the above relationship shows that $\lim_{t \to \infty}P_{k,t} \to P_{k}$ and results into $P_{k} \equiv  0~~ \forall k$, which contradicts the constraint. It means that we do not get a stationary distribution and it is a time dependent satisfying the constraint $\sum_{k=1}^{K_t} P_{k,t} = 1$, with $K_t$ as the maximum degree of the network node, growing at each time step. The network is no longer a scale-free network but converts to a network with uniform degree distribution. We can estimate the value of $K_t$. For large $t$, and for all $k>1$,

\[
P_{k,t} \approx P_{1,t}
\]

Now,
\[
\begin{split}
\sum_{k=1}^{K_t} P_{1,t} = 1\\
K_t = \frac{1}{P_{1,t}}
\end{split}
\]

Thus if $P_{1,t}$ is of order $\frac{1}{t^{\beta-1}}$, $K_t$ is of order $t^{\beta-1}$.

\end{document}